\documentclass[prl,10pt,tightenlines,twocolumn,twoside,showpacs,nofootinbib,superscriptaddress]{revtex4-2}
\usepackage{amsfonts}
\usepackage{float}
\usepackage{placeins}
\usepackage{graphicx}
\usepackage[hidelinks]{hyperref}
\usepackage{color,amsmath,amssymb,bm}
\usepackage{tensor}
\usepackage[normalem]{ulem}  
\usepackage{changepage}   
\usepackage{slashed}
\usepackage{siunitx}
\usepackage{mathbbol}
\usepackage{bbm}
\usepackage{hyperref}
\DeclareSymbolFontAlphabet{\mathbbmm}{bbold}
\DeclareMathAlphabet{\mathbbmsl}{U}{bbm}{m}{sl}
\DeclareSymbolFontAlphabet{\mathbb}{AMSb}%

\usepackage[normalem]{ulem}

\def \no {\nonumber}
\def \a {\alpha}
\def \b {\beta}

\def \d {\delta}
\def \e {\epsilon}
\def \g {\gamma}

\NewCommandCopy{\oldo}{\o}
\DeclareRobustCommand{\o}{\ifmmode\omega\else\oldo\fi}

\def \l {\lambda}
\def \m {\mu}

\def \s {\sigma}

\def \th {\theta}

\def \pd {\partial}

\def \<{\langle}
\def \>{\rangle}
\def \+{\dagger}

\def \[{\left[}
\def \]{\right]}

\def \vp {\bm{p}}

\def \vx {\bm{x}}

\def \va {\bm{a}}

\def \vn {\bm{n}}
\def \vk {\bm{k}}

\def \vL {\bm{L}}
\def \vJ{\bm{J}}

\def \vv {\bm{v}}

\def \ve {\varepsilon}

\def \veta {{\bm{\eta}}}

\def \vd{{\bm{d}}}
\def \vl{{\bm{l}}}

\def \D {\Delta}

\def \<{\langle}
\def \>{\rangle}
\def \+{\dagger}

\def \[{\left[}
\def \]{\right]}

\def \srho 
{{\tilde \rho}}

\renewcommand{\sout}{\bgroup \color{red} \ULdepth=-.5ex \ULset}

\def\blfootnote{\xdef\@thefnmark{}\@footnotetext}

\newcommand{\beq}{\begin{equation}}
\newcommand{\eeq}{\end{equation}}
\newcommand{\bea}{\begin{eqnarray}}
\newcommand{\eea}{\end{eqnarray}}

\newcommand\sect[1]{\textit{#1.}---}

\begin{document}
\title{
	Side jump and anomalous Lorentz transformation of a coherent Dirac wave packet
}
\author{Jin-Rui Wu}
\affiliation{School of Physics\&Electronics, Hunan University, Changsha 410082, China}
\affiliation{Hunan Provincial Key Laboratory of High-Energy Scale Physics
	and Applications, Hunan University, Changsha 410082, China}
\author{Shuai Y.F.~Liu}
\email{lshphy@hnu.edu.cn}
\affiliation{School of Physics\&Electronics, Hunan University, Changsha 410082, China}
\affiliation{Hunan Provincial Key Laboratory of High-Energy Scale Physics
	and Applications, Hunan University, Changsha 410082, China}

\date{\today}

\begin{abstract}
Spin–orbit coupling makes the mean position of a spinning particle transform anomalously under a Lorentz boost, with an additional shift known as a side jump. In this work, we develop a trace method to derive an exact but compact transformation kernel for the probability center of a general three-dimensional Dirac wave packet under an arbitrary finite Lorentz boost. The expression includes the physics of  finite momentum width, spin–momentum correlations, and coherent phase structure without a semi-classical expansion. We identify texture-dependent side jumps even at fixed initial mean momentum, spin, and orbital angular momentum, as well as a non-monotonic mass dependence for Gaussian packets at fixed momentum width. The explicit kernels further resolve the spin and orbital angular-momentum contributions for massive and massless fermions, providing a kinematic basis for covariant particle-based spin transport that retains wave-packet information.
\end{abstract}

\maketitle
\sect{Introduction}
Side jumps connect spin and orbital motion and have attracted broad interest across classical and quantum physics. They arise in the Lorentz transformation of spinning objects~\cite{Pryce:1948pf,Newton:1949cq} and relativistic electron and photon wave packets~\cite{Bliokh:2011fi,Bliokh:2011xa,bliokh2017position,Bliokh:2024awa,Bliokh:2026abs,Smirnova:2017twi}. In condensed matter, side jumps contribute to anomalous Hall transport~\cite{PhysRevB.2.4559,Nagaosa:2009ycg}, with connections to momentum-dependent spin textures, Bloch-state geometry, and spin Hall effects~\cite{PhysRevB.73.075318,RevModPhys.87.1213,Xiao:2009rm}. Related spin-dependent transverse shifts have also been studied in connection with the angular-momentum decomposition and spatial distributions of polarized hadrons~\cite{Burkardt:2002hr,Carlson:2007xd,Lorce:2018zpf,Lorce:2020onh}.
In relativistic heavy-ion collisions, side-jump effects, which play a central role in establishing a consistent kinetic theory of spin transport, have been studied in the context of chiral kinetic theory~\cite{Chen:2014cla,Chen:2015gta,Hidaka:2016yjf,Hattori:2019ahi} and  spin polarization~\cite{Liu:2019krs,Weickgenannt:2020aaf}. These studies, together with other theoretical approaches~\cite{Liang:2004ph,Becattini:2013fla,Karpenko:2016jyx,Li:2017slc,Sun:2017xhx,Pang:2016igs,Liu:2021uhn,Fu:2021pok,Becattini:2021suc,Becattini:2021iol,Fu:2022myl,Wu:2022mkr, Sheng:2019kmk,Muller:2021hpe,Weickgenannt:2022jes,Sheng:2022ffb,Sheng:2022wsy,Sheng:2019kmk,Sheng:2022ffb,Sheng:2022wsy,Li:2022vmb},  contribute to our understanding of the rich polarization phenomena observed in heavy-ion collisions~\cite{STAR:2017ckg,STAR:2018gyt,Niida:2018hfw,Adam:2019srw,ALICE:2021pzu,STAR:2022fan,STAR:2008lcm,ALICE:2019aid,Singha:2020qns,ALICE:2022dyy}.

Recent work has shown that orbital polarization might dominate over spin polarization in strongly magnetized rotating QCD matter~\cite{Fukushima:2024tkz}. This highlights the importance of the orbital structure of fermion wave functions and calls for a covariant description of spin transport that retains both the orbital and spin structure of coherent wave packets. In particular, a kinematic prerequisite for this is understanding their anomalous probability-center shifts and spin–orbital decomposition under general Lorentz transformations. Although coherent wave-packet structure has received less attention in heavy-ion spin transport, it has been extensively studied in photon and electron optics~\cite{Bliokh:2017uvr}. However, explicit boost calculations for structured beams address specific configurations~\cite{Smirnova:2017twi},  while kinetic descriptions typically rely on semiclassical expansions~\cite{Hattori:2019ahi,Weickgenannt:2020aaf}. Explicit, exact kernels for these transformations remain unavailable for general wave packets with finite momentum width and arbitrary spin and phase structure.

In this work and the companion paper~\cite{Li:2020vwh}, we develop a trace method to derive this kernel in an exact but compact form for general three-dimensional Dirac wave packets. We further relate their spin--orbital angular-momentum decomposition to a covariant particle description for massive and massless fermions.

\sect{Theoretical formalism}
To calculate the side jump, we first write the Dirac wave packet following the conventions of the textbook~\cite{Peskin:1995ev}:
\begin{align}
\label{eq:wavepackage}
    &\psi(x) = \int \frac{d^3 \vk}{(2\pi)^3 \sqrt{2 \varepsilon_k}} a(k) u(k) e^{-i(\varepsilon_k t - \vk \cdot \mathbf x)}\;,\\
\label{eq:DiracSpinor}
    &u(k)= S(-\boldsymbol\eta_k) u(0),\; u(0)=\sqrt{m}\left(
    \begin{tabular}{c}
         $\xi$  \\
         $\xi$ 
    \end{tabular}
    \right)
\end{align}
Here, $\eta_k$ is related to $ \vk$ by $ \vk = m\sinh|\eta_k|\hat{\boldsymbol\eta}_k$.  With $ \check{u}=(1,\bm0),h_{\veta}=\Lambda_{\veta/2} \check{u},u_{\bm\eta}=\Lambda_{\veta} \check{u}$, we found a useful gamma-matrix representation of the transformation matrix:
\begin{equation}
	\label{eq:sgamma}
    S(\boldsymbol\eta)= h_{\bm\eta}^{\mu}\gamma_\mu \check{u}^\nu \gamma_\nu,\; h_{\veta}=(u_{\bm \eta}+\check{ u})/\sqrt{2(u_{\bm\eta}\cdot \check{u}+1)}.
\end{equation}
The same matrix $S(\veta)$ boosts a spinor wave packet as
\begin{align}
    \psi^\prime(x)=S(\boldsymbol\eta)\psi\left(\Lambda^{-1}_{\boldsymbol\eta} x\right) 
    \label{eq:psiprime}
\end{align} 
under a boost with rapidity $\veta$, where
\begin{equation}
	\Lambda_{\boldsymbol{\eta}} = \left(
	\begin{array}{cc}
		\cosh \eta & -\sinh \eta \hat{\boldsymbol{\eta}} \\
		-\sinh \eta \hat{\boldsymbol{\eta}} & I + (\cosh \eta - 1)\hat{\boldsymbol{\eta}}\hat{\boldsymbol{\eta}}
	\end{array}
	\right).
\end{equation}
 The spinor $\xi$ defines the spin direction $\hat{\vn}$ as $\xi^\dagger{\bm\sigma}\xi=\hat{\vn}$. $\xi$ and $\hat{\vn}$ may depend on $\vk$, encoding nontrivial spin momentum correlations. The  amplitude $a(k)=a_0(k)e^{i\phi(k)}$ satisfies $\int d^3\vx\,\psi^\dagger\psi=\int d^3\vk/(2\pi)^{3}\,|a(k)|^2=1$.
 
Assuming $a(k)$ vanishes rapidly at infinity, the boosted probability center $\int d^3\vx\; \vx\, \psi^{\prime\dagger}(x)\psi^\prime( x)$ can be expressed in momentum space using $\widetilde\psi^\prime= \int d^3\vx \psi^\prime(x) e^{-i \vk^\prime \cdot  \vx}$:
\begin{align}
\label{eq:x1}
    &\bar{\mathbf x}_{\boldsymbol \eta}(t) = i \int \frac{d^3\mathbf k^\prime}{(2\pi)^3} \widetilde\psi^{\prime\dagger}(\mathbf k^\prime,t) \partial_{\mathbf k^\prime}\widetilde\psi^\prime(\mathbf k^\prime,t)\\
	\label{eq:psikp}
    &\widetilde\psi^\prime(\mathbf k^\prime, t) =  \frac{\varepsilon_k}{\varepsilon_{k^\prime}} \frac{a_0(k)e^{i\phi}}{\sqrt{2\varepsilon_k}}S(\boldsymbol{\eta})S(-\veta_{k}) u(0) e^{-i\varepsilon_{k^\prime} t}.
\end{align}
Here, $k^\prime = \Lambda_{\veta} k$ and $\vk'$ denotes the spatial part of $k'$. Using $d^3 \mathbf k^\prime / (2 \varepsilon_{k^\prime}) = d^3 \mathbf k / (2 \varepsilon_k)$,  we obtain a compact form of Eq. (\ref{eq:x1}) after a nontrivial derivation:
\begin{align}
\label{eq:xbars}
   & \bar{\vx}_{\veta}(t)=   \bar{\vv}_{\veta}t+\vd_{\veta}+\delta \vx_{\veta}\\
   \label{eq:veta}
       &\bar{ \vv}_{\veta} \equiv \int \frac{d^3 \vk}{(2\pi)^3} |a(k)|^2\vv'_{\vk},\;\vv'_{\vk}= \frac{ \vk^\prime}{\varepsilon_{k^\prime}}	\\
	\label{eq:deta}
	&\vd_{\bm \eta} \equiv \int \frac{d^3 \vk}{(2\pi)^3}|a(k)|^2[-\partial_{ \vk^\prime}\phi+i\xi^\dagger\partial_{ \vk'}\xi]\\
     \label{eq:dx}
    &\delta \mathbf x_{\boldsymbol\eta} \equiv \int  \frac{d^3\vk}{(2\pi)^3 2\varepsilon_{k^\prime}}
    |a(k)|^2 \Big[\sinh \eta (\hat{\mathbf n}\times\hat{\boldsymbol\eta})\nonumber\\
    &\qquad-\frac{\cosh \eta}{\varepsilon_k+m}(\hat{\vn}\times\vk)+\frac{\cosh \eta-1}{\varepsilon_k+m}(\hat{\vn}\times \vk)\cdot\hat{\veta} \hat{\veta}\Big]
\end{align}
The $\bar{\vv}_\veta$, $\vd_\veta$ and $\d \vx_\veta$ terms arise from $\vk'$ derivatives acting on $e^{-i \ve_{k'}t}$, \{$e^{i\phi}$, $u(0)$\} and $ S(-\veta_k)$, respectively. $\vd_\veta$ combines the phase gradient and the Berry connection $i\xi^\dagger\partial_{ \vk'}\xi$ in a gauge-invariant form. The derivation repeatedly uses the identity $u^{\dagger}_s(k)S^2(\eta) u_{s'}(k)=2\varepsilon_{k'}\delta_{ss'}$.

Defining $\langle\cdots\rangle=\int d^3\vk/(2\pi)^3\,|a(k)|^2(\cdots)$ and denoting the Pauli--Lubanski vector of each momentum mode in the original frame by $W^\mu_{\vk}$~\cite{Lubanski:1942A,Lubanski:1942B,Weinberg:1995mt} (see long paper~\cite{Li:2020vwh}), we write $\bar{\vx}_{\veta}=\<\vx_{\vk}\>$ with a physically intuitive kernel whose structure can even be guessed on physical grounds:
\begin{align}
\label{eq:xintgrand}
&\hspace{-0.5em}\vx^{i}_{\veta,\vk}=[\frac{\sinh\eta}{\ve_k\ve_k'} \bm{W}_{\vk}\times\hat{\veta}]^i+\Lambda^i_\veta{}_{j}\vx_{\bm0\vk}^j-\vv^{\prime i}_{\vk} \d t'_{\vk}+\vv^{\prime i}_{\vk}t\\
&\hspace{-0.5em}W^\mu_{\vk} = \frac 1 2 (\hat{\vn}\cdot \vk, m\hat{\vn} + \frac{(\hat{\vn} \cdot \vk)\vk}{\ve_k+m}),\\
&\hspace{-0.5em}\vx_{\bm0\vk}=\frac{-\hat{\vn}\times\vk}{2\ve_k(m+\ve_k)}-\pd_{\vk}\phi+i\xi^{\dagger}\pd_{\vk}\xi, \d t_{\vk}'=\Lambda^0_\veta{}_{j}\vx_{\bm0\vk}^j
\end{align}
However, to our knowledge, this is the first exact derivation of this kernel from Dirac theory for general three-dimensional wave packets. The first term has the known classical side-jump form but follows from an exact quantum derivation; the remaining terms describe the boosted initial position, equal-time adjustment, and free propagation. This kernel gives the exact centroid transformation for fully 3D Dirac wave packets, enabling the 3D analysis explicitly called for in Ref.~\cite{Smirnova:2017twi}.
The distinction between energy and charge centers has long been discussed~\cite{Bliokh:2011xa,bliokh2017position,Bliokh:2024awa,Bliokh:2026abs}. Here, their comparison reduces to evaluating $\<\ve_{k'}\vx_{\vk}\>/\<\ve_{k'}\>$ and $\<\vx_{\vk}\>$ with the same explicit kernel.

Evaluating $\delta\vx_{\veta}$ is indeed cumbersome, so we developed nontrivial Mathematica code in our companion paper's initial version~\cite{Li:2020vwh}, reducing several pages to one line. However, as a derivation tractable by hand would be of broader interest, we now develop a general and transparent \textbf{trace method}, detailed in the companion paper. Since the imaginary contributions ultimately cancel, we obtain $\delta\vx_{\veta}$ in Eq.~(\ref{eq:dx}) from the real part:
\begin{align}
	\label{eq:deltax}
	&\delta\mathbf{ x}_{\bm \eta}
	=  \text{Re}\Big[ i\int \frac{d^3\mathbf k}{(2\pi)^3} \frac{|a(k)|^2}{2\varepsilon_k'}X(k)\Big] \\
	\label{eq:tracetech}
	&X(k)=\frac{m}{2}\text{Tr}\left\{
	(1+\slashed{\check{u}}\gamma_5\slashed{n})\slashed{h}_{\bm\eta_{k}}\slashed{h}_{\bm\eta}\slashed{h}_{-\bm\eta} \partial_{\mathbf k^\prime}\slashed{h}_{-\bm\eta_{k}}\right\},
\end{align}
\begin{align}
	\label{second_item}
	&\text{Re}[i X(k)]=i \frac{m}{2}\text{Tr}(\gamma^\nu\gamma^\rho\gamma^\sigma
	\gamma^\alpha\gamma^\beta\gamma^\mu\gamma^5)\no\\
	&\qquad\qquad\times{\check{u}}_\mu n_\nu h_{\bm\eta_k,\rho}h_{\bm\eta,\sigma}h_{-\bm\eta,\alpha}\partial_{\mathbf{k}^\prime} h_{-\bm\eta_k,\beta}.
\end{align}
where we express $S(\veta)$ and $S(-\veta_{k})$ with Eq.~(\ref{eq:sgamma}) and use $u(0)u^\dagger(0)=m(1+\slashed{\check{u}})(1+\slashed{\check{u}}\gamma_5\slashed{n})/2 $, with $n^\mu\equiv(0, \hat{\vn })$, and $\slashed{h}_{-\bm\eta_x}\slashed{\check{u}}=\slashed{\check{u}}\slashed{h}_{\veta_x}$. The $\gamma_5$ term alone contributes to $\delta\mathbf{ x}_{\bm \eta}$.

After a basic trace calculation, we obtain
\begin{align}
	\label{eq_trace}
	&\text{Tr}\left(\gamma^\nu\gamma^\rho\gamma^\sigma\gamma^\alpha\gamma^\beta\gamma^\mu\gamma^5\right)=4i\left(\eta^{\alpha\beta}\epsilon^{\mu\nu\rho\sigma}+\eta^{\mu\beta}\epsilon^{\alpha\nu\rho\sigma}\right.\nonumber\\
	&\left.-\eta^{\alpha\mu}\epsilon^{\beta\nu\rho\sigma}-\eta^{\nu\rho}\epsilon^{\sigma\alpha\beta\mu}-\eta^{\sigma\rho}\epsilon^{\nu\alpha\beta\mu}+\eta^{\nu\sigma}\epsilon^{\rho\alpha\beta\mu}\right).
\end{align}
The trace follows from $\g^{\a}\g^{\b}\g^{\m}=\eta^{\a\b}\g^{\m}+\eta^{\b\m}\g^{\a}-\eta^{\a\m}\g^{\b}-i\e^{\l\a\b\m}\g_{\l}\g^5$ and the standard trace formula. Substituting Eq.~(\ref{eq_trace}) into Eq.~(\ref{second_item}) yields
\begin{align}\label{eq:eta epsilon}
	\text{Re}[i X(k)]=
	&-(2m)\left(\eta^{\alpha\beta}\epsilon^{\mu\nu\rho\sigma}+\eta^{\mu\beta}\epsilon^{\alpha\nu\rho\sigma}-\eta^{\alpha\mu}\epsilon^{\beta\nu\rho\sigma}\right.\nonumber\\
	&~\left.	-\eta^{\nu\rho}\epsilon^{\sigma\alpha\beta\mu}-\eta^{\sigma\rho}\epsilon^{\nu\alpha\beta\mu}+\eta^{\nu\sigma}\epsilon^{\rho\alpha\beta\mu}\right)\no\\
	&\times{\check{u}}_\mu n_\nu h_{\bm\eta_k,\rho}h_{\bm\eta,\sigma}
	h_{-\bm\eta,\alpha}\partial_{\mathbf{k}^\prime} h_{-\bm\eta_k,\beta}.
\end{align} 
The explicit form of $h_{\mp\bm\eta_{k}}$ can be expressed as
\begin{align}
	h_{\mp\bm\eta_{k}}^\mu
	=\left(\pm \bar k^\mu+(m+\varepsilon_{ k})\check{u}^\mu\right)/\sqrt{2(\varepsilon_{ k}+m)m},
\end{align}
where $\bar{k}^\mu=\bar{\d}^{\mu\nu}k_\nu\equiv(0,\bm{k})$ with the projector $\bar{\delta}^{\mu\nu}=\eta^{\mu\nu}-\check{u}^\mu\check{u}^\nu$. The derivatives are
\begin{align}
	&\frac{\partial h^\mu_{-\bm\eta_{k}}}{\partial\bar{k}'^i}=\frac{M^{j}_{\;\;i}}{D_{h_k}}\left[ \left( \bar{\delta}^{\mu}_{ j}+\dfrac{\bar{k}^\mu \bar{k}_{j}}{2\varepsilon_k(\varepsilon_k+m)}\right) -\frac{\check{u}^{\mu}\bar{k}_{j}}{2\varepsilon_{k}} \right]
\end{align}
where $M^{j}_{\;\;i}=(\partial \bar k^j)/(\partial \bar k^{\prime i})=[\Lambda_{-\bm\eta}]^{j}_{\;i} -[\Lambda_{-\bm\eta}]^j_{\;0}\bar{k}'_{ i}/\varepsilon_{k'}$ and $D_{h_k}=\sqrt{2m(\ve_k+m)}$.
Despite the frame dependence encoded by $\check{u}$, rotational covariance allows $\delta \vx_{\veta}$ to be expressed through dot and cross products, as in Eq.~(\ref{eq:dx}). The remaining algebra is lengthy but tractable and detailed in our companion paper~\cite{Li:2020vwh}. The method can also be extended to obtain compact Lorentz-transformation formulas for the multipole moments of wave packets.

The side jump is the spatial separation between $\bar{\vx}_\veta(t')$ and the Lorentz-transformed original center trajectory at equal boosted time:
\begin{equation}
\label{eq:transform}
    \left(
    \begin{tabular}{c}
         $t^\prime$  \\
         $\bar{\mathbf x}_{\boldsymbol\eta}(t^\prime)$
    \end{tabular}
    \right)
    =\Lambda_{\veta}
    \left(
    \begin{tabular}{c}
         $t_{*}$  \\
         $\bar{\vx}_{\bm0}(t_{*})$
    \end{tabular}
    \right)+
    \left(\begin{tabular}{c}
    	$0$  \\
    	$\bm{\Delta}$
    \end{tabular}\right),
\end{equation}
The condition $[\Lambda_\veta(t_*,\bar{\vx}_{\bm0}(t_*))]^0=t'$ places the two events at equal boosted time, making $\bm\Delta$ purely spatial~\cite{Chen:2014cla,Chen:2015gta} (see Fig.~\ref{fig:lineshift} in the End Matter). With the choices specified below, Eq.~(\ref{eq:transform}) gives:
\begin{equation}
\label{eq:jump}
    \bm\Delta^i = (\bar{ \vv}^{\prime i} \Lambda^{0}_{\veta~j}-\Lambda^{i}_{\veta~j})\bar{\vx}^j_{\bm0}+\bar{\vx}^i_{\veta}.
\end{equation}
The reference worldline and equal-time hypersurface determine $\bar{\vv}'$: the Lorentz-transformed original center velocity $\bar{\vv}'_o$ for $t'=0$~\cite{Li:2020vwh}, the energy-momentum velocity $\bar{\vv}'_e=\<\vk'\>/\<\ve_{k'}\>$ for parallel reference worldlines, or $\bar{\vv}_\veta$ for $t'=\Lambda^0_\veta{}_{j}\bar{\vx}_{\bm0}^j(0)$, the time of the boosted initial spacetime event. We use the last choice below and note differences where relevant; most examples are unaffected. Inserting the kernel~Eq.~(\ref{eq:xintgrand}) into Eq.~(\ref{eq:jump}) gives:
{
\thickmuskip=4mu
\begin{align}
	\label{eq:jumpfull}
	\setlength{\thinmuskip}{10mu}
	\setlength{\medmuskip}{100mu}
	\setlength{\thickmuskip}{10mu}  
	&\hspace{-0.5em}\bm\D\equiv\bm\D_{\bm w}+\bm\Delta_{\bm v}\equiv  \<\frac{\sinh\eta}{\ve_k\ve_k'} \bm{W}_{\vk}\times\hat{\veta}\>+\<(\bar{\vv}'-\vv'_{\vk}) \d t'_{\vk}\>\\
	&\hspace{-0.5em}\d t'_k=-\sinh\eta(\frac{-\hat{\vn}\times\vk}{2\ve_k(m+\ve_k)}-\pd_{\vk}\phi+i\xi^{\dagger}\pd_{\vk}\xi)\cdot \hat{\veta}
\end{align}
}where $\bm\D_{\bm w}$ averages the known classical side-jump form over momentum modes, while $\bm\D_{\bm v}$ captures correlations between their velocities and boost-induced time shifts. The averaging alone can produce nontrivial finite-width effects; spin textures modify both terms, while phase gradients enter through $\bm\D_{\bm v}$.

\sect{Coherent side-jump examples}
We illustrate how finite momentum width, phase winding, and spin textures modify the side jump using Gaussian wave packets with $|a(k)|^2=\mathcal{N}e^{-\vk^2/(\sigma^2)}$.

We begin with the zero-width limit, where $|a(k)|^2=(2\pi)^3\delta^3(\mathbf k - \mathbf p)$ gives $\D_{\vv}=0$. Equation~(\ref{eq:jump}) then yields $\bm\Delta$ and its 4D form $\Delta^\mu$:
{\thickmuskip=4mu
\begin{align}
	\label{eq:jumpzero}
	\hspace{-0.4em}\Delta^\mu &=\Delta^\mu_{\bm w} =(0,\frac{\sinh\eta}{\ve_p\ve_p'} \bm{W}_{\vp}\times\hat{\veta})^\mu
	= \frac{\varepsilon^{\mu\alpha\beta\gamma} W^\prime_{\vp,\a} u^\prime_\beta \check u_\gamma}{(p^\prime \cdot u^\prime)(p^\prime\cdot\check u)}
\end{align}
}where $\check u=(1,\bm{0})$, $u^\prime \equiv \Lambda \check u$. As only the $\veta$-transverse part contributes, we can replace $W_{\vp}$ by $W^{\prime}_{\vp}=\Lambda_{\veta}W_{\vp}$, which is generally not $W^\nu_{\vp'}$. For massless helicity eigenstates~\cite{Weinberg:1995mt}, this reproduces the known chiral side jump~\cite{Chen:2015gta}.


\begin{figure}[!t]
	\includegraphics[width=0.49\textwidth]{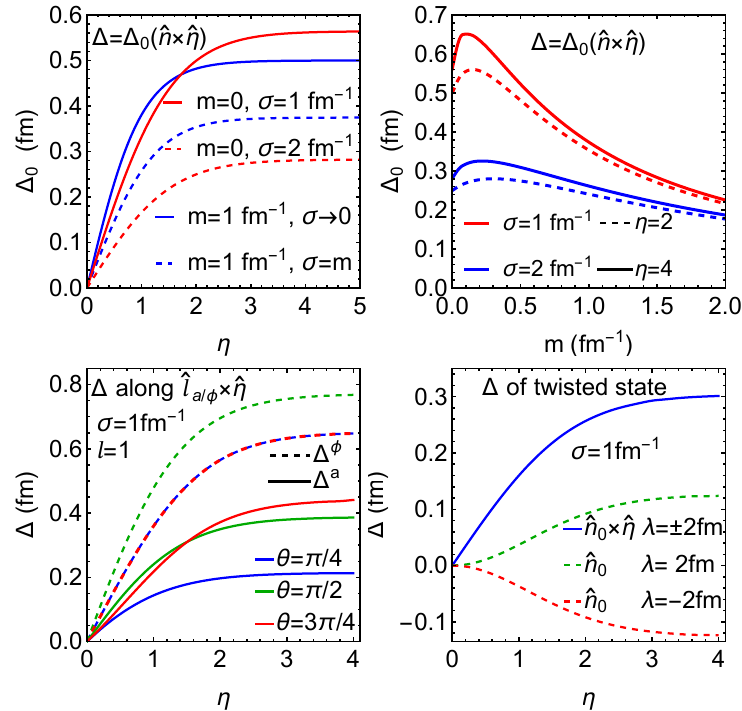}
	\caption{Side jumps of Gaussian packets at rest. (1) and (2): side jumps for fixed spin direction versus $\eta$ and $m$, respectively. (3): helicity ($\Delta^a$) and phase-winding ($\Delta^\phi$) contributions, with $\theta$ the angle between the corresponding axis and the boost. (4):~side-jump components for opposite spin twists. Panels (3) and (4) use $m=0$.
}
	\label{fig:sidejumps}
\end{figure}
For a fixed spin direction, with $\vk$-independent $\hat{\vn}$, $\xi$ and $\phi$,
we have $\bar{\vx}_{0}=0$, implying $\<\d t'_{\vk}\>=0$ and $\bar{\vv}'\<\d t'_{\vk}\>=0$. After symmetry cancellations in $\mathbf k$-space,  the correlation term still contributes to the fixed-$\hat{\vn}$ side-jump~$\bm\D^{\vn}$:
\begin{align}
    &\bm\D^{\vn}\equiv \bm\D_{\bm w}+\bm\D^{\vx}_{\vv},\; \bm\D^{\vx}_{\vv}=\<\frac{\sinh \eta}{2\ve_k\ve_{k^\prime}}\frac{\vk^2- (\vk\cdot\hat{\veta})^2}{2(m+\ve_k)}\Big\>\hat{ \vn}\times\hat{\veta}\no\\
        \label{eq:jumpnfixall}
    &\bm\D^{\vn}=\Big\<\frac{\sinh \eta}{2\ve_k\ve_{k'}}\Big(m+\frac{\vk^2- (\vk\cdot\hat{\veta})^2}{(m+\ve_k)}\Big)\Big\>\hat{\vn}\times\hat{\veta}
\end{align}
The correlation term $\bm\D_{\bm v}$ supplies half of the $\vk^2-(\vk\cdot\hat{\veta})^2$ contribution to $\bm\D^{\vn}$. Both terms contribute to the same transverse structure and must be retained in the exact probability-center shift. 
Dimensional analysis shows that the side jump scales as $1/m$ at fixed $\sigma/m$. The first panel of Fig.~\ref{fig:sidejumps} still shows a $\tanh\eta$-like rise to saturation, while larger widths suppress the side jump. The massless limit remains finite at nonzero width: a packet with $m=0$ and $\sigma=1\,\mathrm{fm}^{-1}$ produces a side jump comparable to that of a zero-width packet with $m=1\,\mathrm{fm}^{-1}$. Thus, momentum width can set the side-jump scale, much as mass does in the zero-width limit. Unlike the $1/m$ scaling in the zero-width limit, the side jump at fixed $\sigma$ depends nonmonotonically on mass, rising from its finite massless value to a maximum before decreasing (second panel). This nonmonotonicity already arises from averaging the classical side jump in $\bm\D_{\bm w}$; the correlation term $\bm\D_{\bm v}$ enhances the shift and moves the peak to smaller masses.

For helicity eigenstates and phase-winding states, orbital angular momentum lies along $\hat{\vl}_{\va}$ and $\hat{\vl}_{\bm\phi}$, with a nonzero Berry connection and phase gradient:
\begin{align}
\hspace{-0.5em}i\xi^{\dagger}\pd_{\vk}\xi
=\frac{-\hat{\vl}_{\va}\times \vk}{2|\vk|(|\vk|+\vk\cdot \hat{\vl_{\va}})}, -\pd_{\vk}\varphi
=l\frac{-\hat{\vl}_{\bm\phi}\times \vk}{\vk^2-(\vk\cdot \hat{\vl}_{\bm\phi})^2}
\end{align}
Here, $\varphi=l\varphi_{\vk}$, with $\varphi_{\vk}$ the azimuthal angle about $\hat{\vl}_{\bm\phi}$; both expressions generalize the usual $z$-axis forms to arbitrary axes. Such phase winding also appears in Landau wave functions~\cite{Xu:2022hql,Chen:2017xrj}, providing a connection to strong-field physics relevant to heavy-ion collisions. For massless packets with $\hat{\vn}=\hat{\vk}$, the side jump arises entirely from these contributions. Exploiting even--odd symmetry in momentum space yields:
{\thickmuskip=4mu
\begin{align}
&\hspace{-0.5em}\bm\D^{\bm a}_{\vv}=
	\Big\<\frac{\sinh \eta}{2\ve_{k'}}\frac{[\vk\cdot(\hat{\vl}_{\va}\times\hat{\veta})]^2}{|\vk|(|\vk|+\vk\cdot \hat{\vl}_{\va})[1-(\hat{\veta}\cdot\hat{\vl}_{\va})^2]}\Big\>		\hat{\vl}_{\va}\times\hat{\veta}\\
&\hspace{-0.5em}\bm\D^{\bm \phi}_{\vv}=l\Big\<\frac{\sinh \eta}{\ve_{k'}}\frac{[\vk\cdot(\hat{\vl}_{\bm \phi}\times\hat{\veta})]^2}{[\vk^2-(\vk\cdot \hat{\vl}_{\bm\phi})^2][1-(\hat{\veta}\cdot\hat{\vl}_{\bm\phi})^2]}\Big\>		\hat{\vl}_{\bm\phi}\times\hat{\veta}
\end{align} 
}These probability-center shifts deviate from the $\sin\th$ dependence of the standard energy-center form $(\tanh\eta/M)\vJ_0\times\hat{\veta}$~\cite{Bliokh:2011xa,Lorce:2018zpf} and encode more information about the microscopic states. In particular, the helicity contribution $\bm\D^{\bm a}_{\vv}$ breaks the symmetry about $\pi/2$, unlike the winding contribution for states with the same $\hat{\vJ}$. Weighting the full kernel by $\ve_k'$ and normalizing by $\<\ve_k'\>$ recovers this energy-center form as expected, with no additional angular dependence. 


For twisted states, motivated by spin precession in momentum-correlated magnetic fields~\cite{PhysRevLett.97.236601,PhysRevB.73.155328}, we take $\xi=e^{i\varphi_{\vk}\hat{\bm t}\cdot\bm\s/2}\xi_0$, with $\varphi_{\vk}=\lambda\vk\cdot\hat{\bm t}$ and $\xi_0$ polarized along $\hat{\vn}^{\bm0}$. Choosing $\hat{\bm t}$, $\hat{\veta}$, and $\hat{\vn}^{\bm0}$ mutually perpendicular makes the Berry connection vanish. Substituting $\hat{\vn}=\xi^\dagger\bm\s\xi$ into Eq.~(\ref{eq:jumpfull}) gives:
\begin{align}
	\label{eq:twist}
	\bm\D^{\bm t}=&\Big\<\frac{\sinh\eta}{2\ve_{k}\ve_{k'}}\cos(\varphi_{\vk})(m+\frac{\vk^2-(\vk\cdot\hat{\veta})^2}{\ve_{k}+m} ) \Big\>(\hat{\vn}^{\bm 0}\times\hat{\veta})\no\\
	&+\Big\<\frac{\sinh\eta}{2\ve_{k}\ve_{k'}}\sin(\varphi_{\vk})\frac{(\vk\cdot \hat{\bm t})(\hat{\veta}\cdot\vk)}{\ve_{k}+m}\Big\>\hat{\vn}^{\bm 0}
\end{align}
The first term in Eq.~(\ref{eq:twist}) shares the usual transverse structure with a $\cos\varphi_{\vk}$ weight. The second adds a shift along the initial spin axis $\hat{\vn}^{\bm0}$ that reverses with $\lambda$. Opposite twists therefore give different side jumps despite identical initial mean momentum, energy, spin, and total angular momentum (fourth panel of Fig.~\ref{fig:sidejumps}).
The kernel explicitly shows that this anomalous spin-parallel shift vanishes for the energy center, since its energy-weighted integrand is odd in $\hat{\veta}\cdot\vk$.

\sect{Decomposition of J and implication to HIC}The spin--orbital decomposition of angular momentum is of broad interest across physics~\cite{Ji:1996ek,Bliokh:2017uvr,Lorce:2018zpf,Speranza:2020ilk}. In chiral transport, an ``orbital contribution" to the polarization is needed to recover the thermal limit~\cite{Chen:2014cla,Chen:2015gta,Liu:2019krs}. In our previous transport study~\cite{Liu:2019krs}, this contribution enters through gradients of the phase-space distribution, complicating a direct particle-based evaluation of momentum-resolved polarization. We therefore examine the decomposition of $J$ at the wave-packet level that fits the ``particle picture''-based spin transport model, further motivated by the dominance of orbital contributions in strongly magnetized rotating matter~\cite{Fukushima:2024tkz}.

Using the kernel in Eq.~(\ref{eq:xintgrand}) and defining $\D_L(\vk)=x_{\veta,\vk}-\Lambda\bar{x}_{\bm0}$, we obtain an explicit decomposition $J^{\mu\nu}=S^{\mu\nu}+L^{\mu\nu}$ consistent with its Lorentz covariance:
{\thickmuskip=3.5mu
\begin{align}
\label{eq:jdecomp1}
&\hspace{-0.5em}S^{\mu\nu}=\big\< \varepsilon^{\mu\nu\a \l } \frac{ W_{\a}'\check{u}_\l}{k'\cdot \check{u}}\big\>,\; L^{\mu\nu}=\<\D_L^\mu k^{\prime\nu}-\D_L^\nu k^{\prime\mu}\>\\
\label{eq:jdecomp2}
&\hspace{-0.5em}\D_L^\mu=\D_{\bm w}^\mu(\vk)+\Lambda^{\mu}_{\veta}{}_j (\vx^j_{\bm0\vk}-\bar{\vx}^j_{\bm0})- \frac{k^{\prime\mu}}{k'\cdot\check{u}}(\d t_k'-\<\d t'_k\>)\\
\label{eq:jdecomp3}
&\hspace{-0.5em}\D_{\bm w}^\mu(\vk)=\frac{\varepsilon^{\mu\alpha\beta\gamma} W'_{\vk,\a} u^\prime_\beta \check u_\gamma}{(k^\prime \cdot u^\prime)(k^\prime\cdot\check u)},\;\d t'_k=\check{u}_\a\Lambda^{\a}_{\veta}{}_j \vx^j_{\bm0\vk}
\end{align}
}Here, $\check u$ makes the frame dependence explicit. The third term in Eq.~(\ref{eq:jdecomp2}) cancels in Eq.~(\ref{eq:jdecomp1}), leaving only the side-jump and position--momentum correlation terms.

\begin{figure}[!h]
	\includegraphics[width=0.49\textwidth]{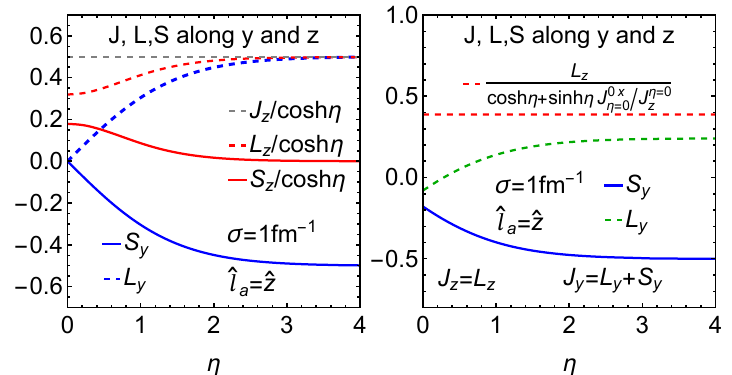}
	\caption{
Spin--orbital decomposition of Gaussian chiral wave packets under boosts along $+y$, with initial mean momenta $\vk_0=(0,0,0.5)\,\mathrm{fm^{-1}}$ (left) and $(0,-0.5,0)\,\mathrm{fm^{-1}}$ (right).
}
	\label{fig:jpls}
\end{figure}

Figure~\ref{fig:jpls} shows the redistribution between spin and orbital angular momentum under boosts, while $J^{\mu\nu}$ transforms covariantly. For the left panel, this behavior follows analytically from the kernels $S=(0,k'_y/k',k_z/k')/2$, $L_w=(0,-k_y'/k',\cosh\eta k_z/k-k_z/k')/2$, and $L_A=(0,0,\cosh\eta-\cosh\eta k_z/k)/2$, after using the symmetries of the momentum integrals.



For Gaussian chiral packets $|a|^2\propto e^{-(\vk-\vk_0)^2/(\s^2)}$, the decomposition following Eq.~(\ref{eq:jdecomp1}) can be expressed as
\begin{align}
	\label{eq:chiralLS}
	&\hspace{-0.5em}\bm S=\frac{\hat{\vk}_0}{2}\big\<\frac{\mathrm{k}_0}{\mathrm{k}}\big\>
	-\frac{\s^2}{4}\big\<\frac{\vk}{k^3}\big\>,\vL
	=\frac{\s^2}{4}\big\<\frac{\vk}{k^3}\big\>+H\,\hat{\vl}_{\va},\\
	&\hspace{-0.5em}\vJ=\frac{\hat{\vk}_0}{2}\big\<\frac{|\vk_0|}{|\vk|}\big\>+H\hat{\vl}_{\va},\;H=\pi\s^2\<
	\d(\vk_\perp)\Theta(-\vk\cdot\hat{\vl}_{\va})\>
\end{align}
Integration by parts separates the Berry monopole and Dirac string contributions to $\vL$, with $\vk_\perp$ transverse to $\hat{\vl}_{\va}$ (see appendix). For isotropic massive packets at rest with fixed spin direction, we obtain
\begin{align}
	&\bm S_{\veta=0}=\frac{1+2\<m/\ve_k\>}{6}\hat{\vn},\;	\vL_{\veta=0}=\frac{1-\<m/\ve_k\>}{3}\hat{\vn},
	\\
	&\vJ^{\text{massive}}_{\veta}
	=\frac{1}{2M}\left[	\bar{\ve}'\hat{\vn}	-\frac{(\hat{\vn}\cdot\bar{\vk}')\bar{\vk}'}{\bar{\ve}'+M}
	\right],\; M\equiv\< \ve_k\>
	\\
	\label{eq:chiralspinboost}
	&\vJ^{\text{chiral}}_{\veta}=(1/2M)[\bar{\ve}'\hat{\vl}_{a}-(\hat{\vl}_{a}\cdot\bar{\vk}')\bar{\vk}'/(\bar{\ve}'+M)]
\end{align}
The same boost law holds for the isotropic chiral packet at rest with $\hat{\vn}\to\hat{\vl}_{\va}$ and $M=\<|\vk|\>$. 

Equations~(\ref{eq:chiralspinboost}) and~(\ref{eq:chiralLS}) show that even wave packets of massless particles possess an independent vector-like degree of freedom $\hat{\vl}_{\va}$, which plays a role analogous to that of $\hat{\vn}$ in the massive case, unifying the description of intrinsic angular momentum.
This provides a particle-based picture of the orbital angular momentum of chiral particles discussed in Refs.~\cite{Chen:2014cla,Chen:2015gta,Liu:2019krs}.
Shifting the reference points gives
\begin{align}
	J^{\mu\nu}_{\text{new}}=J^{\m\nu}-\<\bm \d X^\mu k^{\prime\nu}-\bm \d X^\nu k^{\prime\mu}\>
\end{align}

Here $\d X^\mu=\D^\mu+u^\mu\d t/(u\cdot\check u)$ combines a spatial shift with a displacement along the chosen worldline with $u^\mu$. A common shift changes the packet's reference point; $\vk$-dependent shifts change the reference points mode by mode. The kernel identifies these contributions explicitly.



\sect{Summary}
\label{sec_sum}
We have developed a trace method and used it to obtain an exact, compact transformation kernel for general three-dimensional Dirac wave packets under finite Lorentz boosts. Including finite momentum width, coherent phases, and spin textures, the kernel unifies the treatment of probability and energy centers and spin-orbital angular momentum. Its explicit form reveals how wave-packet structure modifies the classical picture of side jumps and anomalous Lorentz transformations.

For Gaussian packets, momentum width plays a role similar to mass, setting a finite side-jump scale even for massless particles. At fixed width, the mass dependence is nonmonotonic, unlike the classical expectation. Helicity and phase-winding states exhibit probability-center shifts with angular dependences beyond the standard scaling with total angular momentum. Opposite spin twists generate opposite spin-parallel shifts despite identical initial mean momentum, energy, spin, and total angular momentum. The kernel explicitly shows that energy weighting restores the standard angular dependence and removes the spin-parallel shifts. It also provides an explicit covariant angular-momentum decomposition and a particle-based picture of orbital angular momentum, with an independent orbital orientation for chiral packets playing a role like spin orientation in massive packets.

These results provide two kinematic ingredients for particle-based spin transport: consistent boosts between two-particle center-of-momentum and laboratory frames, including side jumps and conversions between charge- and energy-center descriptions; and a spin--orbital decomposition compatible with Lorentz covariance and the particle picture. Together, they lay a foundation for implementing total angular-momentum conservation while treating massive and massless fermions on equal footing. Implementing these results would allow us to develop a consistent spin-transport framework that can be further extended to include quantum coherence encoded in wave packets.
\appendix
\section{Additional plots and derivations}
\begin{figure}[!htb]
	\includegraphics[width=0.28\textwidth]{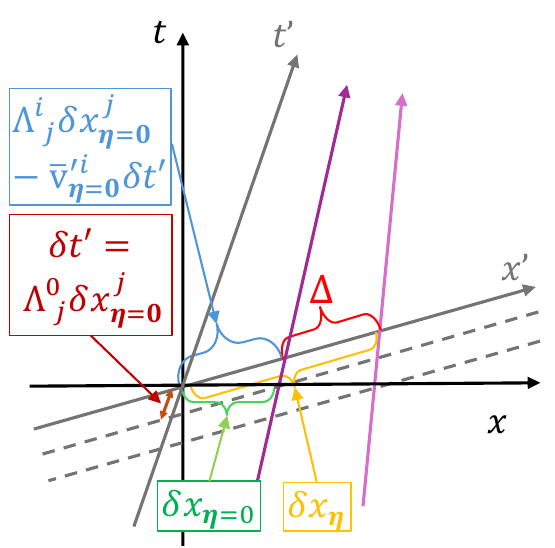}
	\caption{
		Schematic illustration of Eqs.~(\ref{eq:transform}) and (\ref{eq:jump}) . Black and gray axes denote the original and boosted frames. The dark purple line shows the Lorentz-transformed original probability-center trajectory; the light purple line shows the  center trajectory evaluated in the boosted frame. Their spatial separation on $t'=0$ defines the side jump ($\bm\Delta$). For finite-width wave packets, the two trajectories need not be parallel. The same construction applies when $\vd$ is included. 
	}
	\label{fig:lineshift}
\end{figure}
Deriving Eq.~(\ref{eq:chiralLS}) requires care with singularities, as briefly discussed below. Choose $\hat{\vl}_{\va}=\hat{\bm z}$ and write $k=|\vk|$, $\vk_\perp=\vk-k_z\hat{\bm z}$, with $\pd_z\equiv\partial/\partial k_z$ and $\pd_\perp$ the transverse momentum gradient. Then
\begin{align}
	\bm A\equiv i\xi^\dagger\pd_{\vk}\xi=-f\,\hat{\bm z}\times\frac{\vk_\perp}{k_\perp^2}, f=\frac12\left(1-\frac{k_z}{k}\right).
\end{align}
Using the identity $\pd_\perp\cdot(\vk_\perp/k_\perp^2)=2\pi\d^{(2)}(\vk_\perp)$, we obtain
\begin{align}
	\pd_z\left(f\frac{\vk_\perp}{k_\perp^2}\right)
	=-\frac{\vk_\perp}{2k^3},	\pd_\perp\cdot\left(f\frac{\vk_\perp}{k_\perp^2}\right)
	=\frac{k_z}{2k^3}+2\pi f\,\d^{(2)}(\vk_\perp).\no
\end{align}
With the delta function, $k=|k_z|$, $f=\Theta(-k_z)$. The curl thus becomes
\begin{align}
	\pd_{\vk}\times\bm A
	&=\pd_z\left(f\frac{\vk_\perp}{k_\perp^2}\right)
	-\hat{\bm z}\,\pd_\perp\cdot
	\left(f\frac{\vk_\perp}{k_\perp^2}\right)\nonumber\\
	&=-\frac{\vk}{2k^3}
	-2\pi\hat{\bm z}\,\d^{(2)}(\vk_\perp)\Theta(-k_z),
\end{align}
recovering the monopole and Dirac-string terms~\cite{Heras:2018uub}.


\section{acknowledgment}
Jin-Rui Wu and Shuai Liu are supported by NSFC No. 12205090 and Fundamental Research Funds for the central Universities. 

\bibliography{ref}
\end{document}